\documentclass[12pt,preprint]{aastex}
\def\NH{\ifmmode N_{\rm H}\else$N_{\rm H}$\fi}
\def\NHlit{\ifmmode N_{\rm H,lit.}\else$N_{\rm H,lit.}$\fi}
\begin{document}

\title{Extinction Columns and Intrinsic X-ray Spectra of the Anomalous
  X-ray Pulsars}
\author{Martin Durant and Marten H. van Kerkwijk}
\affil{Department of Astronomy and Astrophysics, University of
  Toronto\\  60 St. George St, Toronto, ON\\ M5S 3H8, Canada }

\keywords{pulsars: anomalous X-ray pulsars, X-ray spectroscopy, ISM,
 pulsars: individual (4U 0142+61, 1E 2259+586, 1E 1048.1$-$5937, 1RXS
 J170849.0$-$400910)}

\begin{abstract}
The X-ray spectra of Anomalous X-ray Pulsars have long been fit by
smooth, empirical models such as the sum of a black-body plus a power
law.  These reproduce the $\sim\!0.5$ to 10\,keV range well, but fail
at lower and higher energies, grossly over-predicting the optical and
under-predicting the hard X-ray emission.  A poorly constrained source
of uncertainty in determining the true, intrinsic spectra, in
particular at lower energies, is the amount of interstellar
extinction.  In previous studies, extinction column densities with
small statistical errors were derived as part of the fits of the
spectra to simple continuum models.  Different choices of model,
however, each produced statistically acceptable fits, but a wide range
of columns.  Here, we attempt to measure the interstellar extinction
in a model-independent way, using individual absorption edges of the
elements O, Fe, Ne, Mg and Si in X-ray grating spectra taken with {\em
XMM-Newton}.  We find that our inferred equivalent hydrogen column
density \NH\ for 4U 0142+61 is a factor of 1.4 lower than the
typically quoted value from black-body plus power-law fits, and is now
consistent with estimates based on the dust scattering halo and visual
extinction.  For three other sources, we find column densities
consistent with earlier estimates.  We use our measurements to recover
the intrinsic spectra of the AXPs empirically, without making
assumptions on what the intrinsic spectral shapes ought to be.  We
find that the power-law components that
dominate at higher energies do not extend below the thermal peak.
\end{abstract}
\maketitle

\section{Introduction}
The Anomalous X-ray Pulsars (AXPs) are a group of about six young,
isolated neutron stars showing pulsations with periods of the order
10\,s, whose high-energy luminosity vastly exceeds what is available
in rotational energy losses. They are modeled as {\em magnetars},
their energetics dominated by energy released from a decaying
super-strong magnetic field, $\sim\!10^{14}\,$G externally. See Woods
\& Thompson (2004) for a summary of current observations and their
interpretation in the context of the magnetar model.

Since their discovery, the X-ray spectra of the AXPs have been fitted
with simple continuum models, most commonly the sum of a black body,
representing the peak of the spectrum around a few keV, and a
power law to account for the emission at higher energies.  The overall
spectra are relatively soft, as indicated by high power-law indices.
In the fitting process, the interstellar extinction is estimated as
well, by including a multiplicative term to take account of the
absorption of all elements, parameterized by the hydrogen column
density, \NH.

While these smooth continuum spectral models reproduce the spectra of
all AXP well in the range observed by satellites such as {\em
Chandra}, {\em XMM} and {\em ASCA} (roughly 0.5 to 10\,keV), their
extrapolation fails at both higher and lower energies.  At energies
from tens of keV to $\sim\!1\,$MeV, Kuiper et al.\ (2004) and Den
Hartog (2006) found from RXTE/HEXE
and Integral observations that much harder power-law components are
present, which dominate the total energetics.  In the optical and
infrared, the first detections of AXPs made by Hulleman et al.\ (2000,
2001) already showed that while the optical and infrared emission is
only a tiny part of the energy budget, it is well below (by orders of
magnitude) the extrapolation of the X-ray spectra
dominated by the power law, yet well above the extrapolation of just
a blackbody component.  Thus, the optical and infrared emission also
appears to require a separate emission component.  More recently, the
situation was complicated even further, when Wang et al.\ (2006) found
from mid-infrared observations from {\em Spitzer} that at least one of
the AXPs (4U 0142$+$61) shows evidence of a circumstellar dusty disc.

The above discrepancies led us to wonder whether the measurements of
interstellar extinction were reliable.  These are particularly
sensitive to the the low-energy ($<\!500\,$eV) part of the spectra
(where the absorption is highest), and one would expect that they
would depend strongly on the assumed model for the intrinsic emission.
For instance, the above-mentioned model consisting of a black-body and
a power-law component rises towards lower energies, while a model of,
say, two black-body components, would turn over.  Thus, to match a
given observation, the former would require a larger column density
than the latter.

From previous studies (see Table~\ref{nhlit}), it is indeed clear that
the inferred column density \NH\ depends strongly on the assumed
intrinsic model, with the differences in \NH\ for different models
fitted to the same spectrum far exceeding the statistical uncertainty
obtained for any given one.  Clearly, without knowing the intrinsic
shape of the spectrum, one cannot measure the extinction accurately,
and, conversely, without knowing the extinction independently, one can
obtain only little information about the true intrinsic spectrum.

Another clue that the extinction is not estimated correctly, and hence
that the models used for the intrinsic spectrum are incorrect, comes
from variability studies.  AXP spectra have been seen to vary both
from one epoch to another, and between phase bins (e.g., Woods et al.\
2004; Rea et al.\ 2005), and, generally, different values of \NH\ are
found for these different spectra (see Table~\ref{nhlit}).  Although
an intrinsic, variable contribution to the extinction is not
impossible in light of the discovery of a possible debris disk around
4U~0142+61 (Wang et al.\ 2006), it seems unlikely, especially on the
time-scale of seconds.  Indeed, discrepancies in fitted parameters
obtained from spectra taken with different instruments, which are
often blamed on poor cross-calibration, could well be purely an
artifact of the fitting process, with differences in sensitivities
leading to differences in weight as a function of energy, and hence
different values for the parameters, including the
extinction.\footnote{Another source of differences may be the use of
different sets of cross-sections and abundances; see, e.g., Weisskopf
et al.\ (2004).}

\begin{deluxetable}{lccccccl}
\tablecaption{Hydrogen column densities towards the AXPs inferred
using different models.\label{nhlit}} 
\tablewidth{0pt}
\tabletypesize{\footnotesize}
\tablehead{
\colhead{Object} &\colhead{Telescope/}&
\multicolumn{5}{c}{\dotfill Model\tablenotemark{a}\dotfill} & \colhead{Reference}\\
 & \colhead{instrument}& \colhead{PL} & \colhead{PL+BB} & \colhead{BB+BB} &
\colhead{BR} & \colhead{BR+BB} 
}
\startdata
4U 0142+61 & XMM/EPIC &       \nodata & 0.96 &\nodata & 0.82 &   0.68 &
 G\"ohler et al.\ (2004) \\
 & Chandra/HETGS &            1.43: &   0.88 &\nodata & 0.92 &   0.69 &
 Juett et al.\ (2002)\\
 & ASCA/SIS,GIS &             \nodata & 0.95 &\nodata &\nodata & 0.90 &
 White et al.\ (1996)\\
1E 2259+586&XMM/EPIC &        \nodata & 0.94-1.10\tablenotemark{b} &
                                              \nodata &\nodata &\nodata &
 Woods et al.\ (2004)\\
 & Chandra/ACIS &             \nodata & 0.93 &\nodata &\nodata &\nodata &
 Patel et al.\ (2001)\\
 & ASCA/PSPC &                1.14: &   0.85 & 0.63 &   0.8 &   \nodata &
 Rho \& Petre (1997)\\
1E 1048.1$-$5937&XMM/EPIC&    \nodata & 0.95-1.10\tablenotemark{b} &
0.55-0.67\tablenotemark{b} &\nodata &\nodata  & Tiengo et al. (2005)\\
 & BeppoSAX/LECS &            1.54: &   0.45 & 0.14 &  \nodata &\nodata &
  Oosterbroek et al.\ (1998)\\
1RXS J170849.0& BeppoSAX/LECS&\nodata & 1.3-1.7\tablenotemark{c} &
                                              \nodata &\nodata &\nodata &
 Rea et al.\ (2005)\\
\multicolumn{1}{r}{$-$400910}& BeppoSAX/LECS&
          \nodata & 1.1-1.6\tablenotemark{c} &\nodata &\nodata &\nodata &
 Rea et al.\ (2003)
\enddata
\tablecomments{Column Densities are in units (10$^{22}$\,cm$^{-2}$).
  For all models listed, the reduced $\chi^2$ of the fits was less
  than 2 (those marked with colons are relatively poor fits).}
\tablenotetext{a}{Acronyms are PL: Power Law, BB: Black Body,
  BR: Bremsstahlung.}
\tablenotetext{b}{Ranges represent measurements made at different
  epochs.  One does not expect the column density to vary with time
  (see text).}
\tablenotetext{c}{Range represents measurements in different phase
  bins.  Again, one would not expect the column density to vary with
  rotational phase.}
\end{deluxetable}

From a physical perspective, it is not clear why there should be a
power-law component in the spectra of AXPs.  Recent physical models,
such as those presented by Lyutikov \& Gavriil (2006) can account for
a power-law like high-energy tail, but do not predict a soft part.  In
these models, the blackbody surface flux is modified by interactions
in the outer atmosphere or magnetosphere, for example through the
inverse-Compton scattering of photons off high-energy particles.  This
creates an extended high-energy tail but leaves the spectrum at low
energies (the Rayleigh-Jeans side of the blackbody spectrum)
unaffected.

Absorption in the interstellar medium of X-ray photons is primarily by
the photo-electric capture by electrons in inner shells of metals and
helium.  Since the creation of X-ray absorption models (Morrison \&
McCammon 1983; Balucinska-Church \& McCammon 1992), great advances
have been made in our understanding of absorption edge structure and
energies (e.g., Juett et al.\ 2003), largely driven by the improved
resolution and sensitivity of X-ray observatories.  We are also
beginning to understand more about interstellar abundances (e.g.,
Lodders 2003; but see \S5 below).  With these advances, it is
now possible to measure individual elemental absorbing columns to a
source independently, and so recover the intrinsic spectrum without
further assumptions.

In this paper, we measure the extinction to the four best-studied AXPs
by analyzing the individual absorption edges present in the
sensitivity range of XMM/RGS, and use this to derive intrinsic X-ray
spectra for each AXP, as well as to estimate reddenings at optical and
infrared wavelengths.  In \S2, we present the data sets we use and
their reduction.  We describe in \S3 how we infer individual element
column densities, and what atomic data we use.  In \S4, we use
Monte-Carlo simulations to estimate the uncertainties of our
measurements. We present our results
in \S5, and show de-extincted X-ray spectra in \S6. We continue by
deriving optical extinctions for all sources in \S7, and discussing
the overall spectral energy distribution of the best-studied source,
4U 0142+61, in \S8.  We briefly summarize our results and look forward
to future work in \S9.

\section{Data Reduction}

We searched the {\em XMM-Newton} archive for observations of all the
AXPs.  The {\em XMM-Newton} observatory (Jensen, 1999) provides data from
three separate telescopes simultaneously, but in this work we are
concerned with the Reflection Grating Spectrometer (RGS) instruments
(den Herder et al.\ 2001), which provide high-resolution spectra in the range
6--40\,\AA.  We found a number of long observations for the four
brightest AXPs\footnote{For the AXP 1E 1841-045, two short XMM
observations exist, but we found these contained too few counts to
provide reliable measurements.} (see Table \ref{datasets}; we omitted
shorter data sets with few counts).  For all these, RGS is used with
the same instrumental setup, thus ensuring a fair comparison of the
sources.  We also searched for high-resolution spectra taken by {\em
Chandra}, but found only a few observations.  Since these did not
allow a comparison between sources, we decided not to use these in the
present work.

We reduced the raw data (Observation Data Files or ODFs) with the
20041122 version of the analysis software, XMM-SAS, and calibration
files. We used the pipeline products for the data from the European
Photon Imaging Cameras, EPIC (Str\"uder et al.\ 2001; Turner et al.\
2001; these were used only as 
additional information for the RGS reduction, see below).

A light curve was produces with 10-s bins from the EPIC data. These
showed periods of high background ({\em flaring}), and such periods
were excised from the RGS analysis using Good Time Intervals
(GTIs). The EPIC data was also used for source selection, using the
automated source detection algorithm. This only works for imaging
modes, but fortunately there are three imaging instruments (EMOS1,
EMOS2, EPN), and out of these only one is required. The co-ordinates
of the brightest source were used to extract the RGS spectra. This
eliminates any uncertainties arising from the telescope pointing and
bore-sight correction, since the relative alignment of the telescopes
and instruments is well known.

Final spectra were obtained by subtracting the background and then
converting to flux using the XMM-SAS task {\tt rgsfluxer}.  For our
spectral modeling, we decided to bin to a relatively low resolution
(0.1\,\AA), since the errors associated with bins with few or no
counts is uncertain. We kept a large number of bins in the spectra and
response matrix functions until the final fluxing, as recommended in
the documentation.  The documentation also states that {\tt rgsfluxer}
fluxes are not recommended for final scientific analysis, but for the
relatively low resolution we use and the relatively poor
signal-to-noise ratio of our data, the accuracy is ample.

In order to improve the signal-to-noise ratio, we decided to merge the
spectra from different observations of each object into averaged
spectra.  Here, we must raise two caveats.  The first is that some
AXPs -- 1E 1048.1$-$5937 in particular -- are variable, and the
spectral shape may be different in each observation.  This, however,
should not change our column estimates, since we fit in small spectral
regions around each absorption edge.  The second caveat is that the
amount of extinction to the continuum source might vary if some of it
is intrinsic.  This seems unlikely, but is perhaps not impossible
given the discovery by Wang et al.\ (2006) of a likely debris disc
around 4U 0142$+$61 (although from broad-band observations there has
been no clear evidence for changes in extinction).  Unfortunately, our
individual datasets do not have enough signal to verify this.

Finally, we note that for 1E 2259+586 there is enhanced background
emission due to the supernova remnant surrounding the AXP.  From the
pipeline-produced {\em order images}, however, we find that the
spectrum of the central source remains distinguishable and that there
are no significant background lines which might contaminate it. The
increased background does, however, lead to poorer signal-to-noise
where the AXP flux is low.

\begin{deluxetable}{lcccc}
\tablecaption{Data sets used.\label{datasets}}
\tablewidth{0pt}
\tablehead{
\colhead{Object} &\colhead{Dataset\tablenotemark{a}}& \colhead{Date} &
 \colhead{Exp.\tablenotemark{b}} &
  \colhead{Counts\tablenotemark{c}}\\ 
&&&\colhead{(ks)}&\colhead{(1000)}}
\startdata
4U 0142$+$61 & 0206670101 & 2004-03-01 & 44.1 & 98.5\\
 & 0206670201 & 2004-07-25 &  23.9 & 62.6\\
1E 2259$+$586 & 0038140101 & 2002-06-11 & 52.5 & 20.2\\
 & 0155350301 & 2002-06-21 & 29.0 & 22.8\\
1E 1048.1$-$5937 & 0147860101 & 2003-06-16 & 69.0 & \phn2.3\\
 & 0164570301 & 2004-07-08 & 33.9 & \phn4.6\\
 & 0307410201 & 2005-06-16 & 23.3 & \phn3.4\\
 & 0307410301 & 2005-06-28 & 25.9 & \phn2.6\\
1RXS J170849.0$-$400910 & 0148690101 & 2003-08-28 & 44.9 & 11.2
\enddata
\tablenotetext{a}{XMM Science Archive identifier}
\tablenotetext{b}{Exposure time for the RGS instruments only;
  typically shorter than the total observation time.}
\tablenotetext{c}{In both RGS instruments, in two orders, after
  good-time-interval filtering and background subtraction.}
\end{deluxetable}

\section{Analysis}\label{ana}

For our measurements, we assume that is is possible to find small
regions of a spectrum around an 
absorption edge, over which the intrinsic spectrum is continuous and
well-described by a power-law.
To each spectral region of interest, we fit a function of the form
\begin{equation}
F_\lambda = A \times
\left(\frac{\lambda}{\lambda_{\rm edge}}\right)^\alpha \times 
\left\{ \begin{array}{ll}
1 & \mbox{for }\lambda > \lambda_{\rm edge}\\
\exp \left[-\left(\frac{\lambda}{\lambda_{\rm edge}}\right)^{5/2} \times
N\sigma\right] & \mbox{for }\lambda \leq \lambda_{\rm edge}
\end{array} \right.\label{edgeequation}
\end{equation}
In the fits, the edge wavelength $\lambda_{\rm edge}$ and
photo-ionization cross-section at the edge $\sigma$ were kept fixed,
but the normalization $A$, power-law index $\alpha$ and the column
density $N$ were allowed to vary. Note that the value of the power-law
index which attenuates the cross-section with wavelength here is not
the more familiar 3. The value of 2.5 fits better with theoretical
calculations (e.g., Verner \& Yakovlev 1995).  We note that the fits
below are insensitive to this change, and negligible additional
uncertainty is incurred.

We used theoretical photo-electric cross-sections from Gould \& Jung
(1991), since cross-sections are very hard to measure accurately.
From recent high-resolution X-ray spectroscopy (see below), we know of
several narrow features and additional structure in some of the edges.
In contrast to the overall strength of the ionisation edges, the
strengths of these additional features depend on the degree of
ionisation along the line of sight.  Since we do not have enough
signal to fit for these, we instead mask any affected points.  The
spectral ranges to fit around each edge were chosen by balancing the
requirement of enough data points for robust fits with that of the
intrinsic spectra being well described by power-law form.

For the Oxygen-K edge ($\sigma=5.642\times10^{-19}{\rm\,cm^{-2}}$), we
fitted the range 19--26\,\AA\ and used $\lambda_{\rm edge}=23.1\,$\AA,
as found by Takei et al.\ (2002).  We masked the regions
22.5--23.1\,\AA\ for multiple edges, and 23.25--23.6\,\AA\ for narrow
lines.

For the Iron-L edge(s) ($\sigma=4.936\times10^{-19}{\rm\,cm^{-2}}$),
we fitted the 16--19\,\AA\ range and used an edge wavelength of
17.52\,\AA, following the work of Juett et al.\ (2006, in prep.).
There are no sharp absorption features, but the edge has multiple
components, leading to a complex shape around the edge.  Accordingly,
we masked the range 17.2--17.56\,\AA.

In the case of the Neon-K edge
($\sigma=3.523\times10^{-19}{\rm\,cm^{-2}}$), we fitted the range
13--16\,\AA\ and used $\lambda_{\rm edge}=14.31\,$\AA\ following once
more the work of Juett et al.  Three narrow absorption features fall
into the fitting range, which we excluded by masking
14.44--14.66\,\AA\ and 13.4-13.5\,\AA.

The Magnesium-K edge ($\sigma=2.191\times10^{-19}{\rm\,cm^{-2}}$) is
the only one in our sample which shows no complicated features. We
used $\lambda_{\rm edge}=9.5\,$\AA, as found by Ueda et al. (2005),
and fitted in the range 8.5--10.5\,\AA.

Finally, for the Silicon-K edge
($\sigma=1.476\times10^{-19}{\rm\,cm^{-2}}$), we once more used the
work of Ueda et al., fitting in the range 6.2--7.5\,\AA\ and using
$\lambda_{\rm edge}=6.72\,$\AA. We masked the region
6.61--6.73\,\AA\ for edges of silicon in silicates, which are slightly
shifted from the edge for atoms in isolation.

\section{Fitting Method and Error Determination}\label{errs}
For each edge, we fit the data with the model in
Eq.~\ref{edgeequation}, with $\lambda_{\rm edge}$ and $\sigma$ fixed
to the values given above, and the best values of $A$, $\alpha$, and
$N$ found by $\chi^2$ minimization.  For the uncertainties on each
data point, we use the errors given by {\tt rgsfluxer}.

There are two possible problems with our fitting method.  First, the
uncertainties produced by {\tt rgsfluxer} are known to be poorly
defined for faint sources.  This is because in the high-resolution
input spectra (which are rebinned to our 0.1\,\AA\ bin size in {\tt
rgsfluxer}), bins with zero counts are assigned an arbitrary
uncertainty of 1, when the expectation value for the (faint) source
might genuinely be near zero and hence the associated uncertainty
should be smaller as well.  These assigned errors are propagated, and,
as a result, the uncertainties on the rebinned fluxes are greatly
overestimated, leading to values of reduced $\chi^2$ far smaller than
unity in our fits.  The second possible problem, related to this, is
that for these counting data, it would be more appropriate to use the
Cash statistic rather than $\chi^2$, since the probability
distribution is Poissonian rather than Gaussian.

Fortunately, in practice these problems are not severe.  In a given
fitting region, we find that in our 0.1\,\AA\ bins both the number of
counts and the {\tt rgsfluxer} uncertainty are roughly constant.
Thus, even though the given errors are over-estimated, this effect is
roughly equal for each point and hence it does not effect the best-fit
parameters.  Furthermore, the bins we use are sufficiently wide that
they contain many counts, and hence the assumption of Gaussian
uncertainties is not so bad.  Indeeed, fits made by minimizing the
Cash statistic led to results very close to those found through the
$\chi^2$ method.  We preferred not to use them generally, however,
since with the Cash minimization our fitting routine converged
much less robustly to the global best fit.

In order to estimate the uncertainties on our measurements, we have
conducted Monte-Carlo simulations, in which we produce simulated data
sets which we fit in the same manner as the real data.  We employed
two different methods for simulating the data.  In the first, we used
{\em bootstrapping,} i.e., we produced simulated data sets with the
same number of points as the actual data set by drawing randomly, with
replacement, from the actual $(\lambda,F_\lambda)$ pairs (Press et
al.\ 1992).  In the second method, we simulated data sets based on the
best-fit model, adding Gaussian noise with a variance equal to the
variance of the real data around that model.  Both methods gave
consistent results after 1000 trials and the spread of results in the
one parameter of interest (the column depth), was well-described by a
Gaussian in every case.  The one-sigma confidence region was taken to
be the region enclosed by the 16th and 84th percentile and these are
listed in Table~\ref{columns}.

We note that in some cases, in particular for 1E 1048.1$-$5937, the
uncertainties even for the best measurements are similar to the
measurements themselves, i.e., the measurements of the indivual
columns are not significant.  In evaluating this, however, one should
keep in mind that we know {\em a priori} the locations and shapes of
the absorption edges (from high signal-to-noise measurements of
brighter sources; see \S3).  Being in the Galactic plane, interstellar
extinction is inevitable, and our aim is not to prove the existence of
the features, but just to measure the strength of features known to be
present.  Thus, the measured column densities and associated
uncertainties can be used as given, and, in particular, columns from
different elements can be combined to give a more significant overall
measurement of the extinction.

Finally, possible sources of systematic error ought to be mentioned.
First, we have chosen to fit the spectral sections with a power law.
This should not be a large source of additional uncertainty, since our
wavelength ranges are so small -- $\Delta\lambda/\lambda=0.2\ldots0.3$
-- that a power law should be a good approximation for any smooth
continuum.  Indeed, the fit would be much better constrained if we
could use larger wavelength range, but this would require knowledge of
what the intrinsic spectrum ought to be.  Second, the optical depth in
an edge does not in general have the simplistic form given in
Equation~\ref{edgeequation} - the form is different for every
ionisation stage of every element.  Fortunately, the equivalent width
of the resulting feature -- and thus the total column -- is not
sensitive to such details, since it depends on the total cross
section.  Third, the masking of certain wavelengths due to complicated
near-edge structure is not necessarily the best approach, but the
alternative -- fitting for them -- would not be better, since the
details of the strengths of the lines depends upon the (unknown)
ionization balance along the line of sight; hence, any gain in number
of data points included in the fit would be offset by the required
additional parameters.

\begin{deluxetable}{lccccc}
\tablecaption{Column densities found for each AXP\label{columns}}
\tabletypesize{\footnotesize}
\tablewidth{0pt}
\tablehead{\colhead{AXP}&
\colhead{O K}& \colhead{Fe L\tablenotemark{a}}&
\colhead{Ne K}& \colhead{Mg K}& \colhead{Si K\tablenotemark{a}}\\ 
&\colhead{$(10^{17}{\rm\,cm^{-2}})$}& 
\colhead{$(10^{17}{\rm\,cm^{-2}})$}&
\colhead{$(10^{17}{\rm\,cm^{-2}})$}&
\colhead{$(10^{17}{\rm\,cm^{-2}})$}&
\colhead{$(10^{17}{\rm\,cm^{-2}})$}}
\startdata
4U 0142$+$61 & $28.8\pm4.5$ & $0.7\pm1.4$ & $5.3\pm1.3$ & $2.2\pm0.5$ & $2.0\pm2.7$\\
1E 2259$+$586 &\nodata &$13\pm6$ & $9\pm4$ & $3.6\pm1.4$ & $0.6\pm3.6$ \\
1E 1048.1$-$5937 &\nodata &\nodata & $7\pm7$ & $2.7\pm2.4$ & $11\pm7$ \\
1RXS J170849.0$-$400910 &\nodata &\nodata & $15\pm5$ & $2.9\pm1.8$ & $2\pm5$ 
\enddata
\tablecomments{No value is shown in cases where no reliable fit was
  possible. }
\tablenotetext{a}{We list the results for Fe L and Si K for
  completeness only.  As discussed in the \S4, we believe these are
  less reliable.} 
\end{deluxetable}

\begin{figure}
\begin{center}
\includegraphics[width=0.24\hsize]{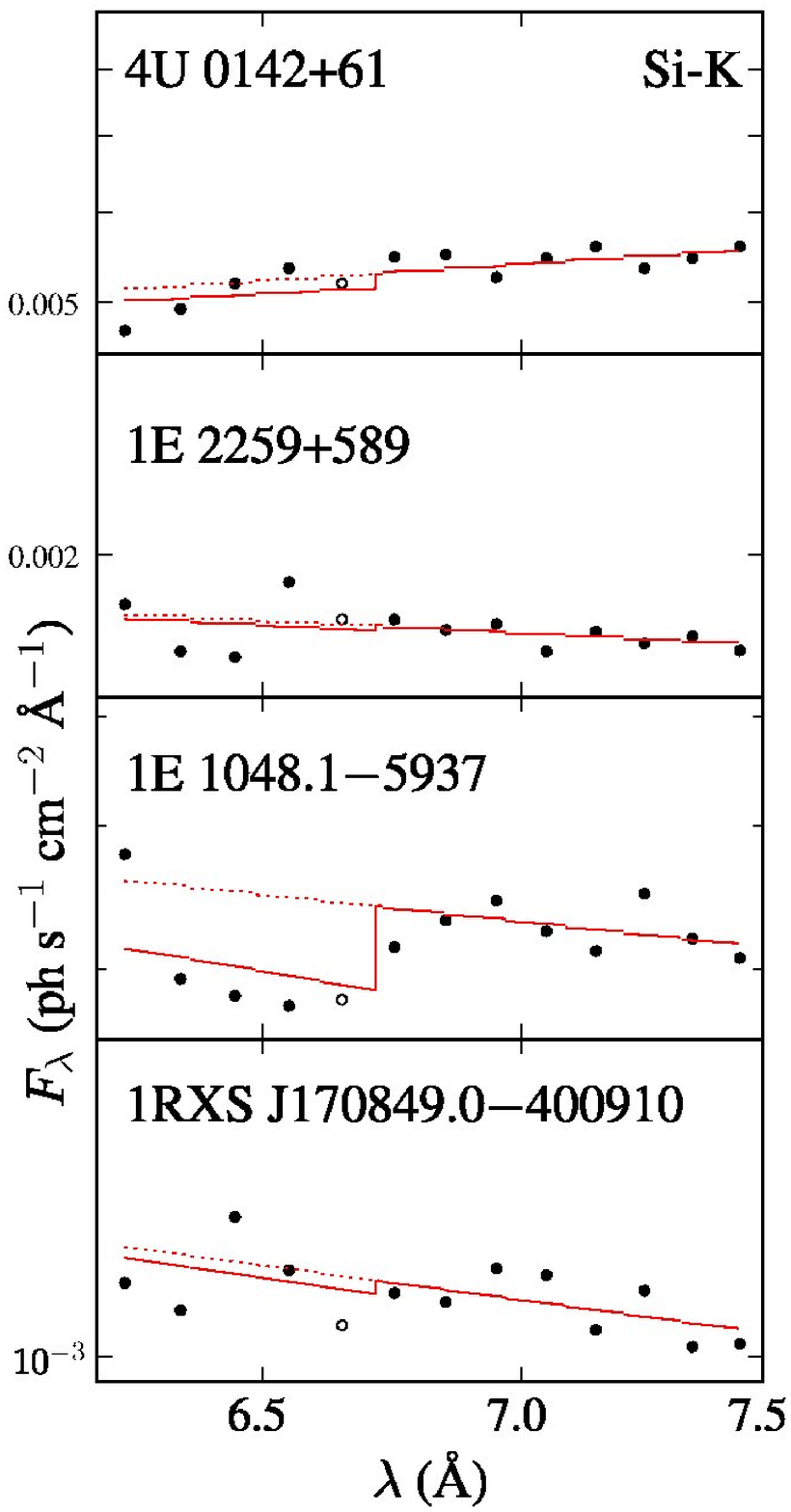}
\includegraphics[width=0.24\hsize]{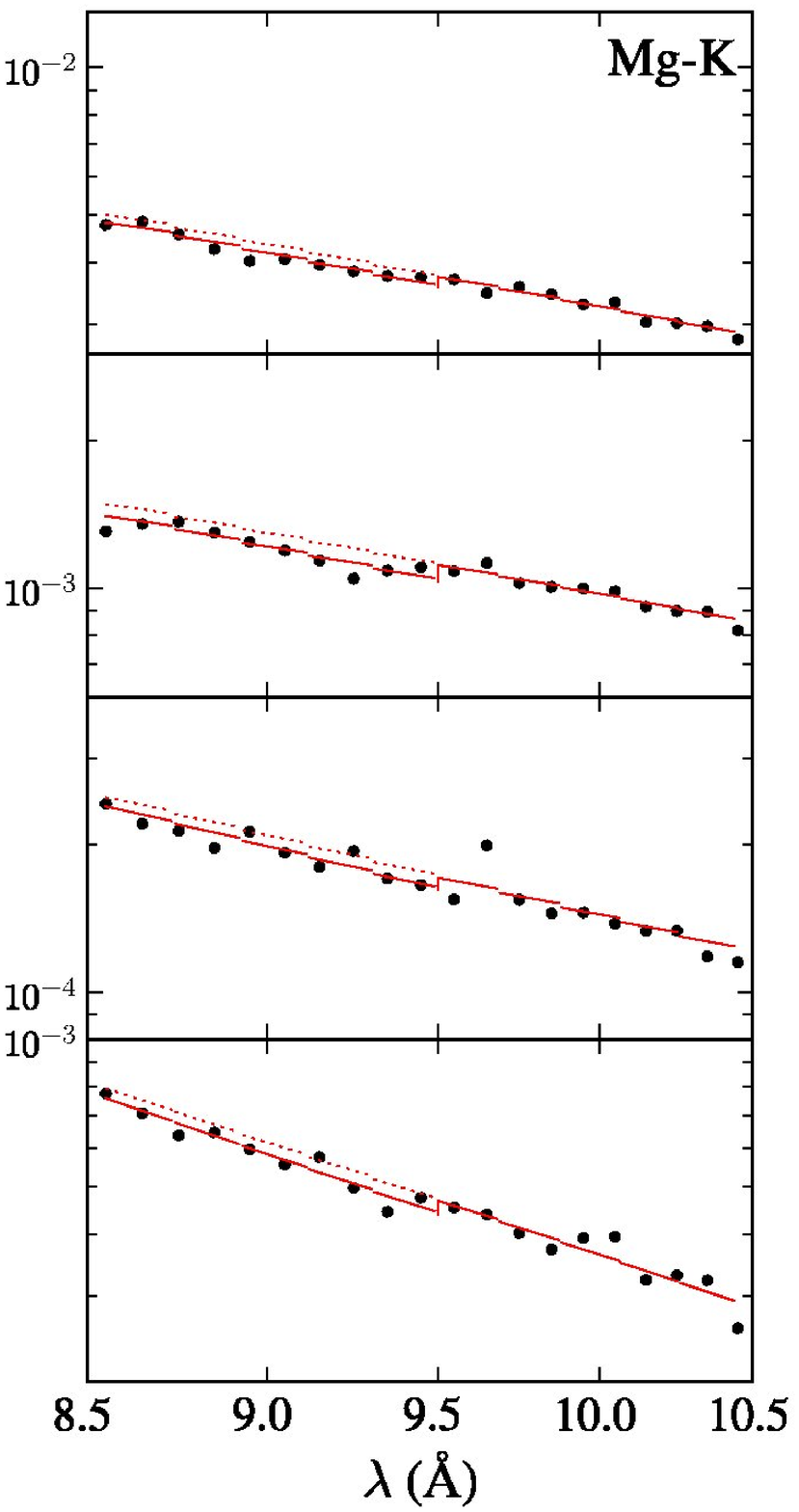}
\includegraphics[width=0.24\hsize]{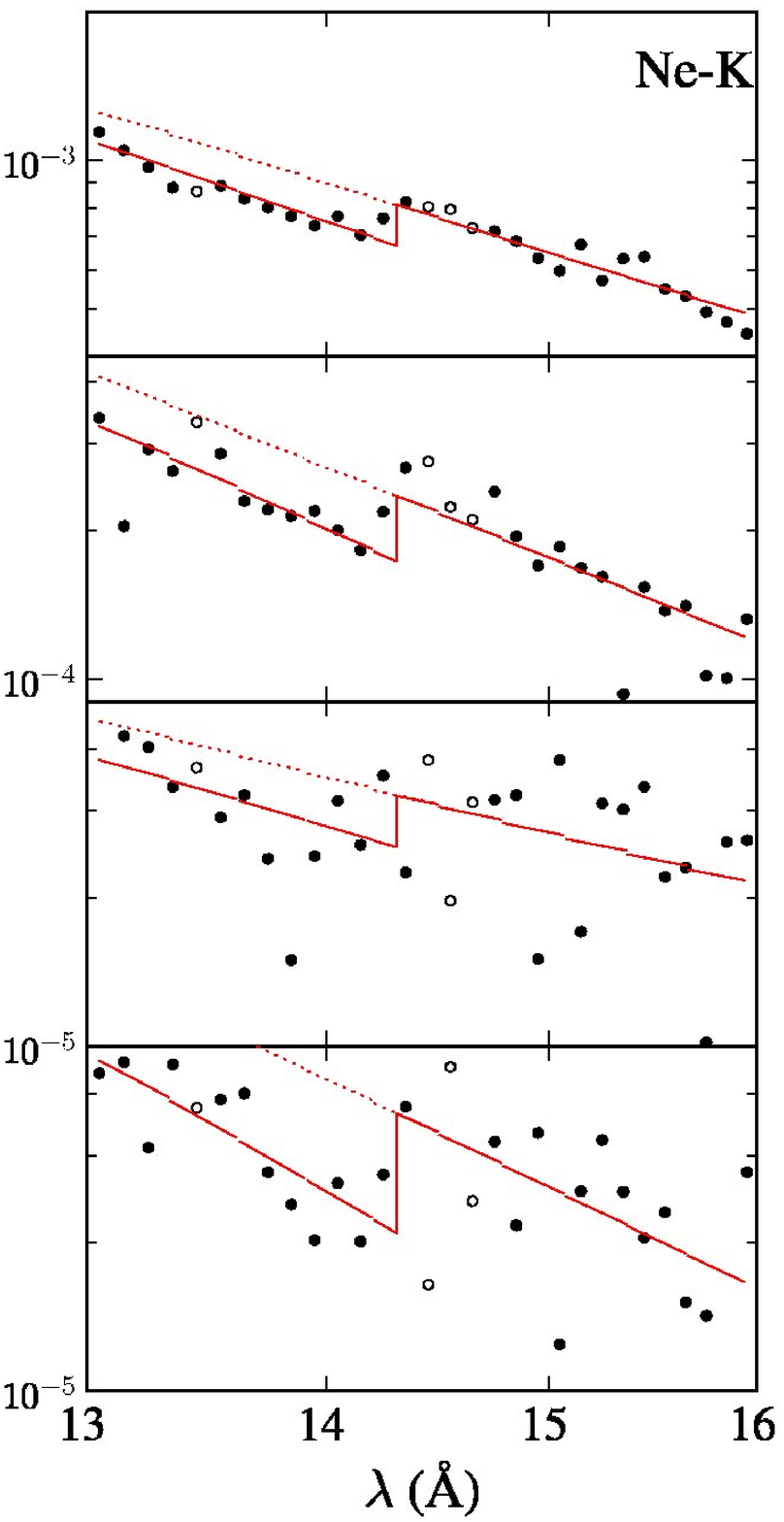}
\includegraphics[width=0.24\hsize]{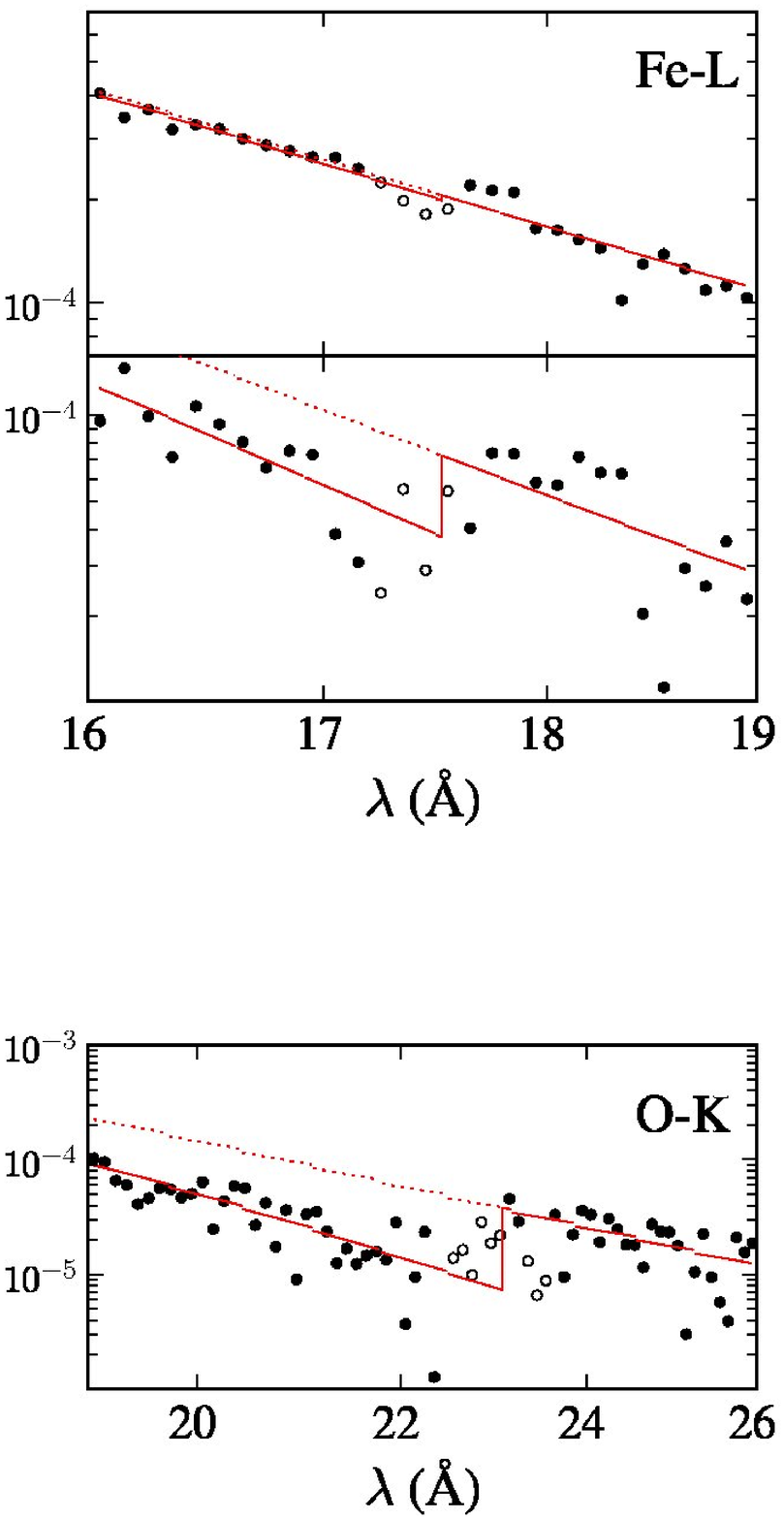}
\caption[0]{Fits obtained for the Si-K, Mg-K, Ne-K and Fe-L edges
  (left to right), with the O-K edge for 4U 0142+61 only in the
  lower-right. Open circles have been 
  excluded from the fit, due to complex edge structures or narrow
  features. For each edge, the scales span the same factor in flux for the
  different AXPs.  As
  discussed in the \S4, we do not use iron or silicon in determining
  the average extinction, since these results are less reliable.
  }\label{FeL}
\end{center} 
\end{figure}

\section{Results}
In Table \ref{columns}, we list the inferred columns for each object,
and in Figure~\ref{FeL}, we show the fits overlaid on the
data points, for those objects where a reasonable fit was possible.
Note that while points with zero or negative flux were included in the
fit, they do not show up in the figures (since the scales are
logarithmic).

From the figures and the table, one sees that the values for iron and
silicon are in every case uncertain, due to the low flux and low
optical depth for the former and the low sensitivity and relatively
poor calibration of the RGS at short wavelengths for the latter.
Since they do not add information, we will not use these results any
further. 

For Neon and Magnesium, we find fair measurements for all sources, and
one sees that their relative abundances are roughly consistent from
one source to another. 1RXS J170849.0$-$400910 appears to be an
exception to this, its magnesium to neon ratio being relatively high,
but with the large uncertainties for this highly reddened source, they
are still consistent within the errors.  From the columns given, it is
immediately clear that 4U 0142$+$61 is least extincted, followed by 1E
1048.1$-$5937, 1E 2259$+$586 and 1RXS J170849.0$-$400910.

We list in Table~\ref{Nh} the implied Hydrogen column densities for
each source, for each reliably measured photo-electric edge.  Here, we
use the abundances of Asplund et al.\ (2004); we discuss this further
below.  Also listed are the weighted means of the column densities.

Comparing our results with those found through broad-band fits using
assumed intrinsic models (Table~\ref{nhlit}), one sees that the values
are consistent with the ranges found previously.  Unfortunately, for
all sources but 4U~0142+61, our uncertainties are too large to
distinguish between different models.  For 4U 0142+61, however, our
value of \NH\ is well determined, and we find it to be lower by a
factor 1.4 than the value inferred from the commonly used black-body
plus power-law model, and closer to that found using the model
consisting of two black bodies.  Interestingly, this is also the
source for which White et al.\ (1996) found a discrepancy between the
extinction and the brightness of its X-ray scattering halo, while
Hulleman et al.\ (2004) found a discrepancy between X-ray extinction,
as inferred from the black-body plus power-law model, and optical
reddening.  With our new value, the different measurements are all
consistent.

In the sections below, we will use our \NH\ values to derive intrinsic
spectra and optical extinctions.  Before doing so, it is worth
stressing that while we quote values of \NH, our measurements are of
the Neon and Magnesium (and Oxygen for 4U 0142+61) column densities.
Thus, the accuracy with which we can determine columns of a given
other element (or dust) depends not only on our statistical
uncertainty, but also on the uncertainty on the abundance of that
element relative to Neon and Magnesium (as well as on possible
corrections for the extent to which a given element is locked up in
optically thick dust; e.g., Wilms et al.\ [2000]; for a general
caution on the implications of the set of abundances used, see
Weisskopf et al.\ [2004]).

In general, the relative abundances of refractive elements are known
precisely from studies of meteorites, but those of the volatiles are
much more uncertain, as has recently become apparent again from the
controversy on the abundances of Oxygen and Neon.  Briefly, Asplund et
al.\ (2004) inferred from improved models for the solar atmosphere
that the solar Oxygen and Carbon abundances were much lower than
thought previously.  The revised abundances, however, led to
discrepancies between helioseismology and models for the solar
interior (e.g., Schmeltz et al.\ 2005).  To remedy this, it has been
suggested that the abundance of Neon might have to be revised upwards
(e.g., Bahcall et al.\ 2005), but this leads to a number of other
problems (e.g., Young 2005; Drake \& Testa 2005).

From our data, an increased Neon abundance seems unlikely: the value
of the Neon to Magnesium ratio we find, ${\rm Ne}/{\rm Mg}=2.4\pm0.7$,
is much closer to the solar-abundance value of 2.0 from Asplund et
al.\ (2004) than the value of 5.8 hypothesised by Bahcall et al.\
(2005) for consistency with helioseismology.  For Oxygen, we have only
one measurement, for 4U 0142+61, which gives ${\rm O}/{\rm
Mg}=13.1\pm3.6$.  This is also much closer to the revised solar
abundance of 13.5 of Asplund et al.\ (2004) than the old one of~22.4
listed by Anders \& Grevesse (1989).\footnote{The good match to the
solar abundances is perhaps surprising, given the discrepant abundance
rations found in other recent X-ray spectroscopic studies (N. Schultz,
2005, personal comm.); it may be a consequence of the fact that AXPs
are not affected by binary interactions.}

Given the above, we are fairly confident that for columns of other
refractive elements (or of dust), any additional uncertainty beyond
the statistical one is small.  However, we are less confident for the
abundances of Hydrogen, Helium, Carbon, Nitrogen, and Oxygen.  This
will affect our corrections for extinction below, especially at low
energies. 

\begin{deluxetable}{lccccc}
\tablecaption{Inferred Hydrogen column densities and amounts of visual
  extinction\label{reds}\label{Nh}} 
\tablewidth{0pt}
\tablehead{\colhead{AXP}& 
\colhead{$\NH({\rm O~K})$}&
\colhead{$\NH({\rm Ne~K})$}&
\colhead{$\NH({\rm Mg~K})$}&
\colhead{$\langle\NH\rangle$}&
\colhead{$A_V$}
}
\startdata
{\em Abundance\tablenotemark{a}}& $4.6\times10^{-4}$&
$6.9\times10^{-5}$& $3.4\times10^{-5}$\\
[1ex]
4U 0142$+$61 & $0.60\pm0.09$& 
  $0.77\pm0.19$& $0.65\pm0.15$& $0.64\pm0.07$& $3.5\pm0.4$\\
1E 2259$+$586&    \nodata & 
  $1.23\pm0.58$& $1.06\pm0.41$& $1.12\pm0.33$& $6.3\pm1.8$\\
1E 1048.1$-$5937 &\nodata & 
  $1.0\pm1.0$&  $0.8\pm0.7$&  $0.87\pm0.57$& $4.9\pm3.2$\\
1RXS J170849.0$-$400910&\nodata&
  $2.1\pm0.7$&   $0.9\pm0.6$&   $1.4\pm0.4$&  $7.7\pm2.2$
\enddata
\tablecomments{All column densities \NH\ are in units of
  $10^{22}{\rm\,cm^{-2}}$.  Column $\langle\NH\rangle$ is the weighted
  mean of all measurements.  For previous determinations of the
  hydrogen column density, from broad-band spectral fits, see
  Table~\ref{nhlit}. $A_V$ is the extinction (in magnitudes) in the
  $V$-band; the errors listed are statistical only, and do not include
  systematic uncertainties in the conversion from~\NH.  The \NH\
  values themselves do not include several possible systematic errors
  (see \S6).}
\tablenotetext{a}{Elemental abundance $N_{\rm O,Ne,Mg}/\NH$ used to
  convert from the measured column from Table~\ref{columns} to the
  equivalent Hydrogen column listed here.  See \S3 and 5 for details.}
\end{deluxetable}

\section{Intrinsic Spectra}

The main question underlying this work is the nature of the intrinsic
spectra of AXPs.  As discussed in \S1, different simple models
reproduce the spectral data equally well, and they do not allow one,
e.g., to distinguish between spectra that rise or fall with decreasing
energy at low energies.  With the column densities obtained above, we
can de-extinct the observations to find the intrinsic spectra
empirically.  For this purpose, rather than attempt to model the
extinction in detail (e.g., Wilms et al.\ 2000), we will use a
simplified model that can be described and reproduced easily; this
will be sufficient to answer the main question, whether the intrinsic
spectra rise or fall at long wavelengths.

For the extinction correction, we convert the average Hydrogen column
densities $\langle\NH\rangle$ from Table~\ref{Nh} to individual
columns for Oxygen, Iron, Neon, Magnesium, and Silicon (using the
abundances of Asplund et al.\ [2004]; see Table~\ref{Nh} for O, Ne,
Mg; furthermore, $N_{\rm Fe}=2.8\times10^{-5}\,\NH$, $N_{\rm Si} =
3.2\times10^{-5}\,\NH$), and correct the spectra for the contribution
of each of these using the simple edge model from Eq.~1 and the
cross sections given following that equation.  Furthermore, we take
into account the absorption by lighter elements, in particular Helium,
Carbon, and Nitrogen, with an additional component
$\exp[-\tau_{25}(\lambda/25\,\hbox{\AA})^\beta]$, where $\tau_{25}$ is
the optical depth at $25\,$\AA.  Inspired by the behaviour of the
total cross-section shown in Fig.~1 of Wilms et al.\ (2000), we choose
$\beta=3$; it is intermediate between the steeper decrease of the
cross-sections of Helium and Hydrogen at these energies and the
shallower one for Carbon and Nitrogen.  For the scaling, we use the
relation given by Morrison \& McGammon (1983),
$\tau_{25}=7.6\times10^{-22}\,\NH$.

The main systematic uncertainty in our correction is due to the
relative abundances, in particular for Oxygen and for the lighter
elements represented by $\tau_{25}$ (see \S5).  The latter uncertainty
could be quite large: for example, calculating $\tau_{25}$ from the
abundances and cross-sections in Morrison \& McCammon (1983), one gets
a correction $\sim$1.7 times larger -- and thus 1.7 times larger
implied intrinsic flux -- than by using those in Wilms et al.\ (2000).
This discrepancy, however, reduces rapidly with decreasing wavelength,
and does not alter the shape of the spectra by much (at least on the
logarithmic scale on which they are shown).

\begin{figure}
\begin{center}
\includegraphics[width=\hsize]{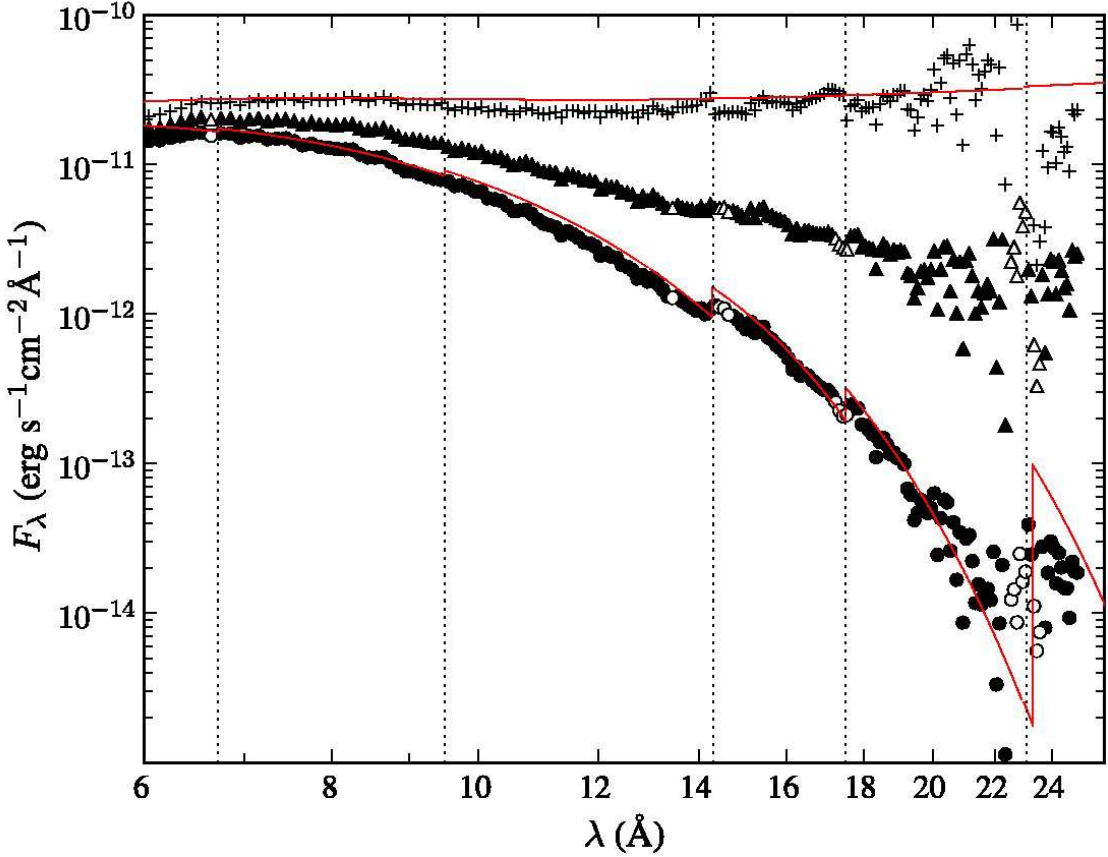}
\caption[0]{Spectra of 4U 0142$+$61 as observed (circles),
  de-extincted with the \NH\ found in this work and the abundances as
  described in the text (triangles) and de-extincted using Morrison \&
  McCammon's cross-sections and $\NH=9.5\times10^{21}{\rm\,cm^{-2}}$
  (crosses).  Open symbols indicate points affected by lines and other
  structure near the different edges.  The over-drawn solid lines are
  the models of White et al.\ (1996) that best fit their broad-band
  {\em ASCA} data (see text).  Vertical lines show the locations of
  the Si-K, Mg-K, Ne-K, Fe-L, and O-K photo-electric absorption
  edges.}\label{deextinct0142}
\end{center} 
\end{figure}

\begin{figure}
\begin{center}
\includegraphics[width=\hsize]{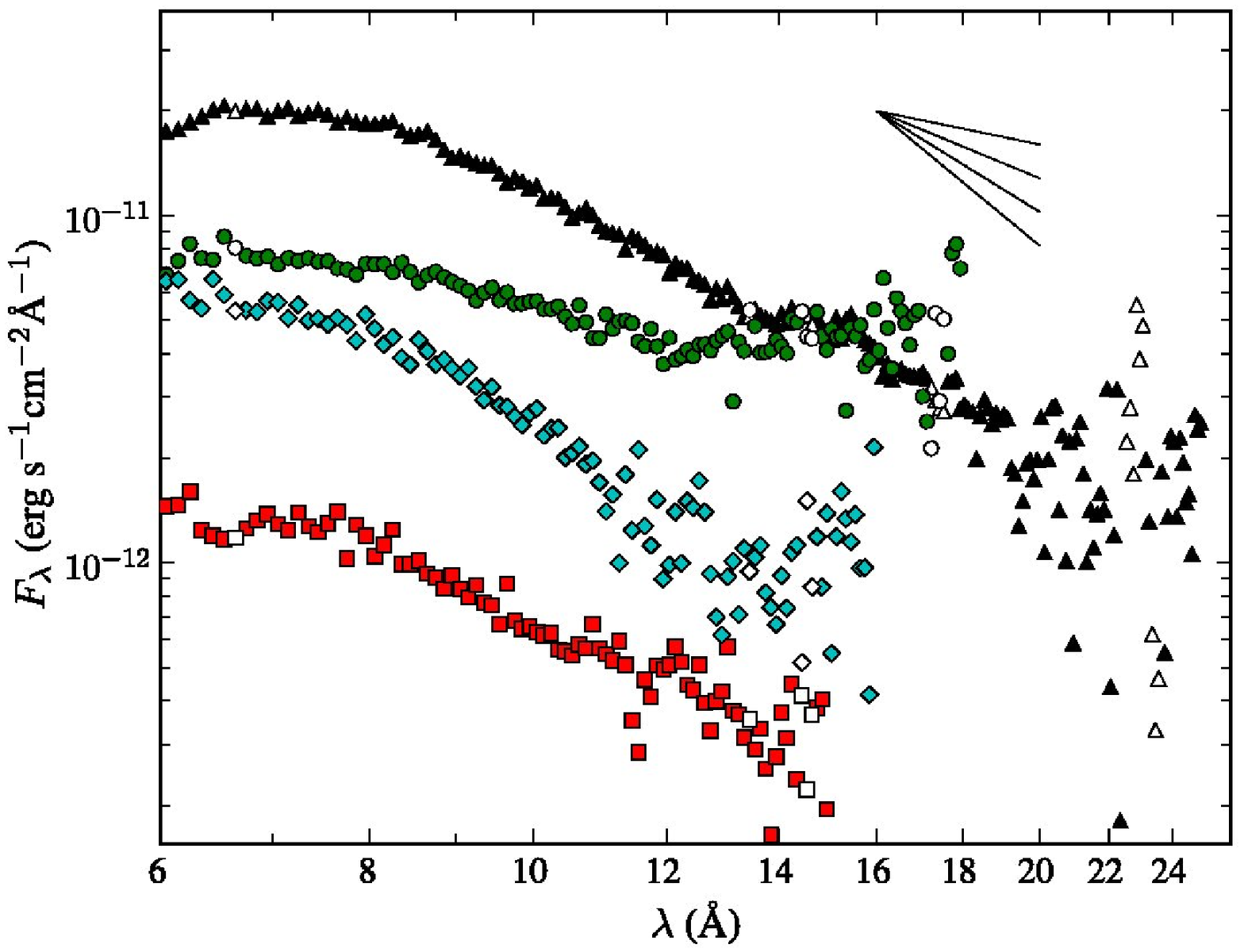}
\caption[0]{Spectra for each AXP, de-extincted with the column
  densities and continuum extinction found in the text.  Black
  triangles are 4U 0142$+$61, green circles 1E 2259$+$586, red squares
  1E 1048.1$-$5937, and cyan diamonds 1RXS J170849.0$-$400910. Open
  symbols are affected by absorption edge structure and absorption
  lines (see text). The spectra have been truncated where the
  signal-to-noise ratio per bin decreases below about 1. Also shown
  are power laws ($F_\lambda\propto \lambda^\alpha$,
  $\alpha=-1,-2,-3,-4$) to guide the eye.}\label{deextinct}
\end{center} 
\end{figure}

In Fig.~\ref{deextinct0142}, we show the observed (circles) and
de-extincted (triangles) spectrum of 4U0142+61: all the edges apparent
in the observed spectrum have been removed in the corrected spectrum
(though discrete features are still evident, such as those due to the
multiple edges of iron at 17.5\,\AA).  From the de-extincted points,
it is clear that the de-extincted spectrum falls with wavelength at
longer wavelengths.

In order to see how it might have been possible to infer a rise at
longer wavelengths from broad-band data, we also show the spectrum
de-extincted with $\NH=9.5\times10^{21}$cm$^{-2}$ (crosses), as found
by White et al.\ (1996) from {\em ASCA} data, and overdraw their fit
of the sum of a power-law and black-body (upper curve) and this same
model extincted by their value of the hydrogen column (lower
curve).\footnote{White et al.\ (1996) used the cross-sections of
Morrison \& McCammon (1983) in calculating extinction, so we have used
this also in this instance only, for consistency.}  One sees from the
Figure that, qualitatively, their fit is not unreasonable, so the
source has not varied much.  Looking in detail, however, especially
near the Oxygen edge, it is clear that their column density is too
high, and that the data are inconsistent with a intrinsic spectrum
rising at long wavelengths.

In Fig~\ref{deextinct}, we show the de-extincted spectra derived for
all four AXPs under consideration.  Two things are immediately
apparent: they are not consistent in shape with one-another, and there
is no continuation at low energies of any power-law component
representing the emission at $>\!2\,$keV: the photon indices, measured
from power-law plus black-body fits, of 2.4 to 4.0 correspond to
slopes of $\alpha=-0.6$, $-0.1$, 0.4, and 1.0
($F_\lambda\propto\lambda^\alpha$), for 1RXS J170849.0$-$400910, 1E
1048.1$-$5937, 4U 0142+61, and 1E 2259+586, respectively, while the
spectra shown have indices of approximately $\alpha=-3$, $-2$, $-2$,
and 0 (if the low-energy tail is taken to be a power-law).  Note that
the largest cause of uncertainty in these derived spectra are the
uncertainties in the column depths of the individual edges, rather
than abundances, detector response or photon statistics.  However, the
conclusion that any power-law component at short wavelengths does not
continue to long wavelength is robust.

The above result raises a paradox: How can it be that even for the
sources other than 4U~0142+61, for which the equivalent hydrogen
columns \NH\ we determine are consistent with those from power-law
plus black-body fits, we do not see the power-law component at long
wavelength? The answer lies in the fact that the low-energy region is
actually not reproduced all that well in typical fits to broad-band
spectra, but this may not be noticed in the formal $\chi^2$ since it
has relatively few counts and thus carries little weight (if it is
included at all).  For a good example of this, see Fig.~2 in Woods at
al.\ (2004): below 0.75\,keV, the data lie systematically $\sim\!20$\%
above the model (particularly easy to see in this figure, since it has
a panel with the ratio of the data to the model, rather than the
usual, less instructive $\chi$ residuals).  With relatively few counts
and larger uncertainties, these points do not affect much the overall
$\chi^2$, but clearly (by eye) they are not well described by the
model (and indeed a bad fit would likely have been found if the data
had been binned more heavily).

While at low energies there is no evidence of continuations of
high-energy power-law components, the spectra also do not decline as
fast as would be expected if they were due to a black-body component.
If the thermal emission arises from the neutron-star surface, as seems
likely, this might simply reflect a range of temperatures on the
surface.  Alternatively, it may indicate that more realistic models
are required to describe the emission, which also include the effects
of magnetic field, interactions with high-energy particles in the
magnetosphere, and gravitational light-bending (the latter
particularly important for phase-resolved spectra).

Finally, looking in detail at the spectra, it appears that for
4U 0142$+$61, there is a hint of a feature in the spectrum at about 13.5\,\AA\
(this is easier to see in Fig.~\ref{SED} at around
$2\times10^{17}$\,Hz).  One could interpret this 
either as a broad absorption feature at $\sim\!13.5\,$\AA, or a broad
emission feature at $\sim\!15\,$\AA.  Assuming it is cyclotron
absorption (emission), i.e., $E_{cyc}=\hbar eB/mc$, this corresponds
to $7.9\times10^{10}\,$G ($7.1\times10^{10}\,$G) for 
electrons or $1.5\times10^{14}\,$G ($1.3\times10^{14}\,$G) for
protons. If the line is
red-shifted, the inferred magnetic field strength would increase by a
factor $1+z_{\rm GR}=(1-2GM/Rc^2)^{-1/2}$, equal to $\sim\!1.3$ at the
surface (for a
neutron star with $M=1.4\,M_\odot$ and $R=10\,$km).  Intriguingly, the
value for proton cyclotron lines is close to the magnetic dipole field
strength inferred from timing measurements, $B_{\rm dip} =
3.2\times10^{19}\surd(P\dot{P}) = 1.3\times10^{14}\,$G (Woods \&
Thompson 2004). 

\section{Optical extinction}

Our revised X-ray extinction measurements also allow us to estimate
the extinction in the optical and infrared.  For this purpose, we use
the relation between the Hydrogen column \NH\ and visual extinction
$A_V$ derived by Predehl \& Schmitt (1995):
$A_V=5.6(\NH/10^{22}{\rm\,cm^{-2}})\,$mag.  The resulting values of
$A_V$ are listed in Table~\ref{reds}.  Here, the errors on $A_V$
listed are statistical, i.e., they do not include the uncertainty in
the conversion factor.  The latter uncertainty could be fairly large,
both because there is substantial scatter in the measurements used by
Predehl \& Schmitt, and because their sample had typically lower
values of extinction.  Furthermore, their hydrogen column is not
measured directly, but is based on X-ray extinction measurements and
thus a measure of elements which contribute significantly to
absorption in X-rays, i.e., Carbon, Oxygen, Neon, and Magnesium.  The
latter should not be a problem in our case, however, since we measure
some of the same elements in Predehl \& Schmitt's sensitivity range
(they did not have the spectral resolution to measure individual
absorption edges), and thus systematic effects in converting to \NH\
should cancel.  The only caveat is that we used the revised solar
abundances of Asplund et al.\ (2004), while Predehl \& Schmitt used,
implicitly, the old solar abundances.  For Neon and Magnesium,
however, the abundances have not changed, while the change for Oxygen
should not have a large effect, since for the one case where we
measure it -- for 4U 0142+61 only -- the inferred \NH\ is consistent
with the values inferred from Neon and Magnesium (see Table~\ref{Nh}).

While the above indicates one should be somewhat careful in using the
absolute values of the reddening, the relative reddenings should be
much more accurate, since any systematic uncertainties in the
conversion should be similar from source to source.  Our results
indicate that in order of increasing reddening, the AXPs are
4U 0142$+$61, 1E 1048.1$-$5937, 1E 2259$+$586 and
1RXS J170849.0$-$400910.  

We can compare our revised estimates with earlier work.  First, for 4U
0142+61, Hulleman et al.\ (2004) found from the colors of field stars,
that the reddening along the line of sight was likely substantially
lower than the value $A_V=5.1$ inferred from the literature values of
\NH\ for the power-law plus black-body model.  Our new value of
$A_V=3.5\pm0.4$ is consistent with the range of $A_V=2$ to 4 seen in
their Fig.~3.  Second, comparing 4U 0142+61 with 1E 1048.1$-$5937,
Durant \& van Kerkwijk (2005) noted that in order for the broad-band
optical/infrared spectrum of these AXPs to have the same shape, the
difference in reddening would have to be $\Delta A_V = 2.5\pm0.5$.
This was inconsistent with previous estimates, which gave very similar
values of $A_V$ (based on the very similar values of \NH; see
Table~\ref{nhlit}), but is consistent with our new results, which give
a difference in reddening of $\Delta A_V = 1.3\pm3$.  Thus, the
optical/infrared spectra of the different AXPs may be similar in
shape, and therefore produced by the same mechanism.

\section{The Spectral Energy Distribution of 4U 0142+61}

With our empirical estimates of the intrinsic (soft) X-ray spectra,
and the revised estimate for optical/infrared reddening, we can
re-examine the spectral energy distributions of the AXPs.  We only
consider 4U 0142$+$61, since this object is the only one for which we
find a significantly different column density from that typically
quoted.  Furthermore, it has the best X-ray data, and the best 
broad-band coverage, from mid-infrared (Wang et al.\ 2006), to
near-infrared and optical (Hulleman et al.\ 2004; Israel et al.\ 2004),
to hard X-rays (Den Hartog et al.\ 2006).  

\begin{figure}
\begin{center}
\includegraphics[width=\hsize]{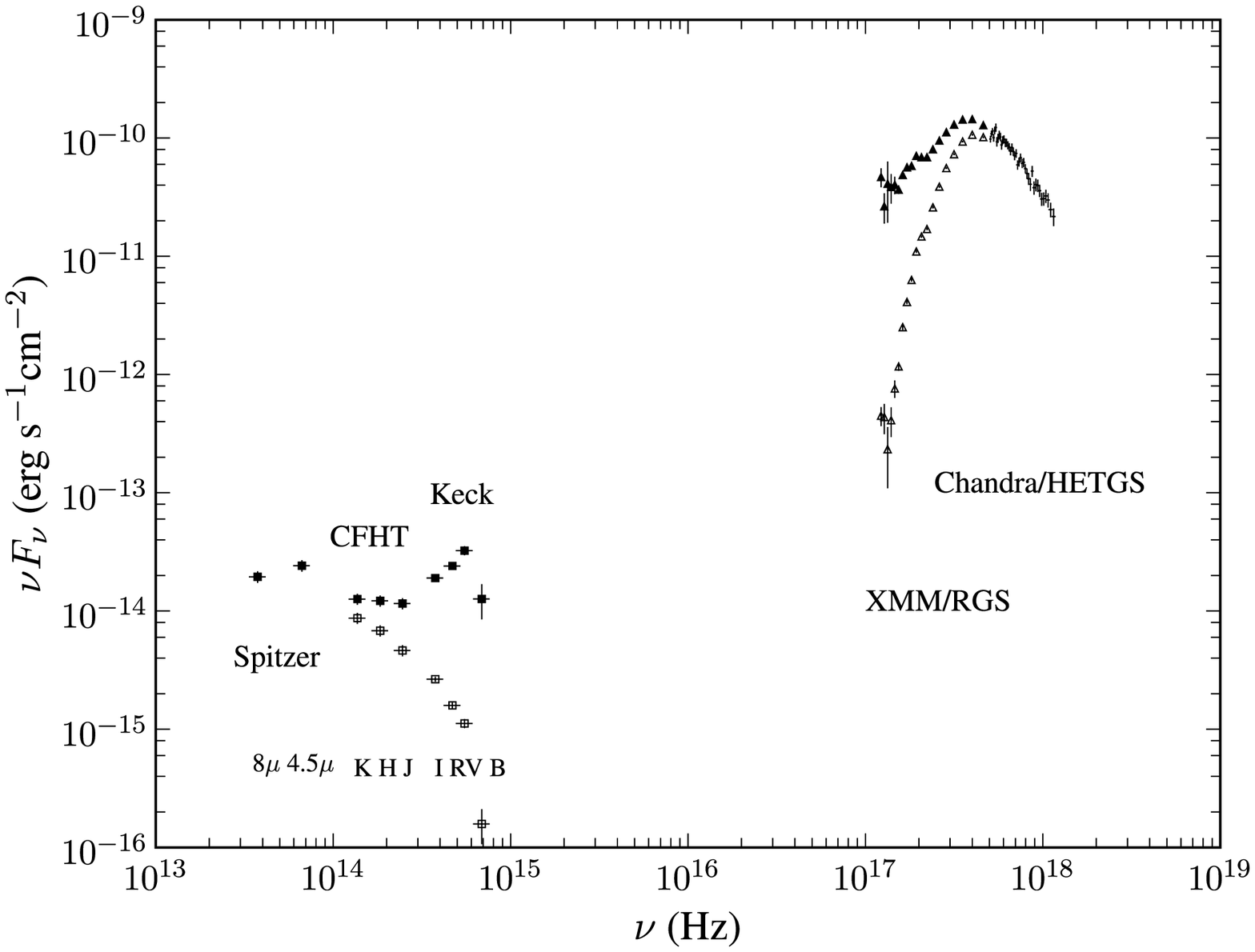}
\caption[0]{Spectral energy distribution for 4U 0142$+$61. Triangles are
XMM/RGS data, as observed (open) and de-extincted (filled) as
described in the text. Open squares are observed broad-band photometry
(mid-IR from Wang et al., 2006; JKH from Israel et al. (2004); BVRI
from Hulleman et al., 2004). Filled squares are the photometric points
de-reddened with $A_V=3.5$ (see text). Crosses in the higher-energy
X-ray part of the spectrum come from {\em Chandra} (Juett et al.,
2002); extinction is not important in this region of the spectrum. XMM
data is binned in frequency ($\delta\nu$) corresponding to
$\delta\lambda = 1$\,\AA, and {\em Chandra} data to 
$\delta\lambda = 0.1$\,\AA\ for clarity.}\label{SED}
\end{center}
\end{figure}

Figure \ref{SED} shows the inferred intrinsic spectral energy
distribution of 4U 0142$+$61, determined with our new values of \NH\
and $A_V$.  The interpretation of the optical and infrared emission is
still unclear: it could be understood as combination of a large
mid-infrared bump combined with a rising power-law that has a sharp
break at B, or as a combination of a much smaller mid-infrared bump
combined with a flatter power-law and a large emission feature
centered between R and~V.  Wang et al.\ (2006) interpret the
mid-infrared bump as arising from a passively illuminated dusty disc.
This disc is presumably formed from supernova fall-back material, and
the data is well fitted by a blackbody from dust at the sublimation
radius.  This disc would also influence the flux in the K-band.  Wang
et al. used our new estimate of \NH, but their results are not very
sensitive to the exact value of \NH, as long as the reddening is above
$A_V\approx2.6$ (Chakrabarty, 2005, pers.\ comm.).

The soft X-ray spectrum in Figure \ref{SED} suggestively has a slope
which would meet up with the optical points if extended. This is in
marked contrast with the earlier power-law models, which grossly
over-predict the optical emission (see \S1). The uncertainty in the
slope of the soft X-rays is dominated by the uncertainty in the
Hydrogen column (and its systematics, particularly due to abundance
uncertainties), so there is not enough information at present to say
whether the break between the B- and V-bands is due to an emission or
an absorption feature.

\section{Conclusions}

We have attempted to measure the extinction to the AXPs without making
assumptions about what their intrinsic spectral shapes might be.  With
our resulting best estimates, we derived intrinsic spectra, which can
be compared with each other as well as with predictions, such as those
from simulations and semi-analytic models that are now being
produced within the magnetar framework (e.g., Lyutikov \& Gavriil
2006; R.\ Fern\'andez \& C.\ Thompson 2006, pers.\ comm.).  

Apart from these comparisons, future work could include extending our
analysis to other sources (once better spectra become available), or
improving the precision of our measurements using further {\em
XMM-Newton} observations (some already taken but not yet public) or
{\em Chandra} spectra (some available).  More interestingly, by
measuring the run of reddening with distance along the line of sight
to the AXPs (using field stars), our extinction estimates can be used
to estimate distances and thus determine the intrinsic luminosities of
the AXPs. It turns out that although the extinctions are not very well
determined, the AXPs fall into areas of rapidly rising extinction
associated with spiral arms, and so can be well localized (Durant \&
Van Kerkwijk, 2006).  Finally, our 
reddening estimates will place on much firmer footing inferences from
further optical and infrared studies, such as could be used, e.g., to
uncover the nature of the break seen between $V$ and $B$ in 4U 0142+61
(Hulleman et al.\ 2004) or to measure the precise parameters of the
possible debris disk around that source (Wang et al.\ 2006).

\acknowledgments We thank Adrienne Juett, Norbet Schultz, and Hermann
Marshall for discussion of the X-ray absorption edges, and Vik Dhillon
for letting us use his {\em XMM-Newton} data on 1E 1048.1$-$5937
before it became public.  We also gratefully acknowledge the comments
of the anonymous referees, which helped us clarify the statistical
significance of our analysis and the presentation of our results.
This work is based on archival observations obtained with XMM-Newton,
an ESA science mission, and made extensive use of NASA's ADS and
Simbad.  We acknowledge financial support from NSERC.


\begin{thebibliography}{}
\bibitem{oldabundence}
Anders, E., Grevesse, N., 1989, Geochimica et Cosmochimica Acta, 53, 197
\bibitem{abundances}
Asplund, M., Grevesse, N., Jacques Sauval, A., 2004, in {\em Cosmic
Abundances as Records of Stellar Evolution and Nucleosynthesis}, ASP
Conference Series, eds F. N. Bash and T. G. Barnes
\bibitem{helioseis}
Bahcall, J. N., Basu, S., Serenelli, A. M., 2005, ApJ, 631, 1281
\bibitem{wabs}
Balucinska-Church, M.,  McCammon, D., 1992, ApJ, 400, 699
\bibitem{dutchy}
 den Hartog, P., Hermsen, W., Kuiper, L., Vink, J., in’t Zand, J., 
\& Collmar, W., 2006, A\&A, accepted, see {\tt astro-ph/0601644}
\bibitem{RGS}
 den Herder, J., et al., 2001, A\&A, 365, L7
\bibitem{neony}
Drake, J., Testa, P., 2005, Nature, 436, 525
\bibitem{me}
 Durant, M., \& van Kerkwijk, M., 2005, ApJ, 627, 376
\bibitem{redclump}
Durant, M., \& van Kerkwijk, M., 2006, ApJ, accepted {\tt (astro-ph/0606027)}
\bibitem{0142models}
G\"ohler, E., Staubert, R., Wilms, J., 2004, MmSAI, 75, 464
\bibitem{energies}
Gould, R. J., Jung, Y., 1991, ApJ, 373, 271
\bibitem{modeconversions}
Ho, W., Lai, D., 2003, MNRAS, 338, 233
\bibitem{H02}
 Hulleman, F., van Kerkwijk, M., \& Kulkarni, S., 2000, Nature, 408, 689
\bibitem{moreHull}
 Hulleman, F., Tennant, A., van Kerkwijk, M., Kulkarni, S.,
 Kouveliotou, C., Patel, S., 2001, ApJ, 563, L49
\bibitem{ferdi}
 Hulleman, F., van Kerkwijk, M., \& Kulkarni, S., 2004, A\&A, 416,
 1037
\bibitem{0142IR}
Israel, G., Stella, L., Covino, S., Campana, S., Angelini, L.,
Mignani, R., Mereghetti, S., Marconi, G., Perna, R., 2004, IAU Symposium no. 218
\bibitem{XMM}
 Jensen, G., 1999, Bulletin of the American Astronomical Society, 32, 724
\bibitem{0142models2}
Juett, A., Marshall, H., Chakrabarty, D., Schulz, N.,  2002, ApJ, 568, L31
\bibitem{xISM}
Juett, A. M., Schulz, N. S., Chakrabarty, D., Canizares, C. R., 2003,
AAS HEAD meeting 7, 06.02
\bibitem{edges2}
Juett et al, 2006, in  preperation
\bibitem{INTEGRAL}
Kuiper, L., Hermsen, W., Mendez, M., 2004, ApJ, 613, 1173
\bibitem{chem1}
Lodders, K., 2003, ApJ, 591, 1220
\bibitem{comp}
Lyutikov, M., \& Gavriil, F., 2006, MNRAS, 368, 690
\bibitem{abundanceratios}
Matteucci, F., Chiappini, C., 2005, PASA, 22, 49
\bibitem{oldsigma}
Morrison, R., McCammon, D., 1983, ApJ, 270, 119
\bibitem{1048models2}
Oosterbroek, T., Parmar, A., Mereghetti, S., Israel, G. L.,
1998, A\&A, 334, 925
\bibitem{2259models}
Patel, S., Kouveliotou, C., Woods, P., Tennant, A., Weisskopf, M.,
Finger, M., G\"o\v{g}\"u\c{s}, E., van der Klis, M., Belloni, T.,
2001, ApJ, 563, L45
\bibitem{PS}
 Predehl, P., \& Schmitt, J., 1995, A\&A, 293, 889
\bibitem{bootstrap}
Press, W., Teukolsky, S. A., Vetterling, W., and Flannery, B., 1992,
{\em Numerical Recipes: the art of scientific computing} 2nd
Edition, Cambridge University Press
\bibitem{1708models2}
Rea, N., Israel, G. L.,  Stella, L., Oosterbroek, T., Mereghetti, S.,
Angelini, L., Campana, S., Covino, S., 2003, ApJ, 586, L65
\bibitem{1708models}
Rea, N., Oosterbroek, T., Zane, S., Turolla, R., M\'endez, M., Israel,
G. L., Stella, L., Haberl, F., 2005, MNRAS, 361, 710
\bibitem{2250models2}
Rho, J., Petre, R., 1997, ApJ, 484, 828
\bibitem{neon}
Schmelz, J., Nasraoui, K., Roames, J., Lippner, L., Garst, J., 2005,
ApJ, 634, L197
\bibitem{EMOS}
Str\"uder, L., et al., 2001, A\&A, 365, L18
\bibitem{edges1}
Takei, Y.,  Fujimoto, R.,  Mitsuda, K.,  Onaka, T., 2002,ApJ, 581, 307 
\bibitem{1048models}
Tiengo, A., Mereghetti, S., Turolla, R., Zane, S., Rea, N., Stella,
L., Israel, G. L.,  2005, A\&A, 437, 997
\bibitem{EPN}
Turner, M., et al., 2001, A\&A, 365, L27
\bibitem{edges3}
Ueda, Y.,  Mitsuda, K., Murakami, H., Matsushita, K., 2005, ApJ, 620,
274
\bibitem{disc}
Wang, Z., Chakrabarty, D., Kaplan, D., 2006, Nature, 440, 772
\bibitem{abundancesaffectNh}
Weisskopf, M., O'Dell, S., Paerels, F., Elsner, R., Becker, W.,
Tennant, A., Swartz, D., 2004, ApJ, 601, 1050
\bibitem{moreedges}
Werner, D., \& Yakovlev, D., 1995, A\&AS, 109, 125
\bibitem{specs}
White, N. E., Angelini, L., Ebisawa, K., Tanaka, Y., Ghosh, P., 1996, ApJ,
463, L83
\bibitem{imms}
Wilms, J.; Allen, A.; McCray, R., 2000, ApJ, 542, 914
\bibitem{wood}
 Woods, P., \& Thompson, C., 2004, in ``Compact stellar X-ray sources'',
 eds Lewin, W., van der Klis, M.
\bibitem{2259xspec}
Woods, P., Kaspi, V., Thompson, C., Gavriil, F., Marshall, H.,
Chakrabarty, D., Flanagan, K., Heyl, J.,  Hernquist, L.,
2004, ApJ, 605, 378
\bibitem{solarneon}
Young, P., 2005,  A\&A, 444, L45
\bibitem{silvie}
Zane, S., Turolla, R., Stella, L., Treves, A., 2001, ApJ, 560, 384 
\end{thebibliography}
\end{document}